\documentclass[aps,twocolumn,superscriptaddress,showpacs]{revtex4}

\usepackage[dvips]{graphicx}  
\usepackage{color}  
\newcommand{\nc}{\newcommand}           
\nc{\vc}[1]     {\mbox{\boldmath $#1$}} 
\nc{\bra}       {\langle}               
\nc{\ket}       {\rangle}               
\nc{\bras}[1]   {\langle#1|}            
\nc{\kets}[1]   {|#1\rangle}            
\nc{\del}       {\partial}              
\nc{\red}[1]    {\textcolor{red}{#1}}  
\nc{\blue}[1]   {\textcolor{blue}{#1}}  
\nc{\green}[1]   {\textcolor{green}{#1}}  
\nc{\beq}     {\begin{eqnarray}}
\nc{\eeq}    {\end{eqnarray}}

\nc{\mydraft}	{\setlength{\topmargin}{-1.0cm}} 
\mydraft

\begin{document}

\title{Tensor-optimized shell model for the Li isotopes \\with a bare nucleon-nucleon interaction}

\author{Takayuki Myo\footnote{myo@ge.oit.ac.jp}}
\affiliation{General Education, Faculty of Engineering, Osaka Institute of Technology, Osaka, Osaka 535-8585, Japan}
\affiliation{Research Center for Nuclear Physics (RCNP), Osaka University, Ibaraki, Osaka 567-0047, Japan}

\author{Atsushi Umeya\footnote{aumeya@nit.ac.jp}}
\affiliation{Human Science and Common Education, Faculty of Engineering, Nippon Institute of Technology, Saitama 345-8501, Japan}

\author{Hiroshi Toki\footnote{toki@rcnp.osaka-u.ac.jp}}
\affiliation{Research Center for Nuclear Physics (RCNP), Osaka University, Ibaraki, Osaka 567-0047, Japan}

\author{Kiyomi Ikeda\footnote{k-ikeda@postman.riken.go.jp}}
\affiliation{RIKEN Nishina Center, Wako, Saitama 351-0198, Japan}

\date{\today}

\begin{abstract}
We study the Li isotopes systematically in terms of the tensor-optimized shell model (TOSM) by using a bare nucleon-nucleon interaction as the AV8$^\prime$ interaction.
The short-range correlation is treated in the unitary correlation operator method (UCOM). 
Using the TOSM+UCOM approach, we investigate the role of the tensor force on each spectrum of the Li isotopes.  
It is found that the tensor force produces quite a characteristic effect on various states  in each spectrum and those spectra are affected considerably by the tensor force.
The energy difference between the spin-orbit partner, the $p_{1/2}$ and $p_{3/2}$ orbits of the last neutron, in $^5$Li is caused by opposite roles of the tensor correlation. 
In $^6$Li, the spin-triplet state in the $LS$ coupling configuration is favored energetically by the tensor force in comparison with $jj$ coupling shell model states.
In $^{7,8,9}$Li, the low-lying states containing extra neutrons in the $p_{3/2}$ orbit are favored energetically due to the large tensor contribution 
to allow the excitation from the $0s$-orbit to the $p_{1/2}$ orbit by the tensor force. Those three nuclei show the $jj$ coupling character in their ground states which is different from $^6$Li.
\end{abstract}



\maketitle 

\section{Introduction}

It is an important subject in nuclear physics to understand the nuclear structure from the viewpoint of the nucleon-nucleon ($NN$) interaction. 
The $NN$ interaction has strong tensor forces at long and intermediate distances caused by the pion exchange and strong central repulsion at short distance caused by the quark dynamics.  
It is important to investigate the nuclear structure by treating these characteristics of the $NN$ interaction~\cite{akaishi86,kamada01,arai11}. 
Recently, it became possible to calculate nuclei up to mass around $A\sim 12$ using a $NN$ interaction with the Green's Function Monte Carlo method~(GFMC) \cite{pieper01, pudliner97}. 
At present, this method requires the extreme computational time to be applied to heavier nuclei. 
It is desired to develop a new method to calculate nuclear structure with large nucleon numbers by taking care of the characteristic features of the $NN$ interaction.

The presence of the tensor force in the $NN$ interaction induces $d$-wave components in a nucleus, in particular, for proton-neutron ($pn$) pairs as the deuteron.  
The $d$-wave component in the deuteron makes the system bound via the large $s$-$d$ coupling of the tensor force. 
This $d$-wave component is also found to be spatially compact as compared with the $s$-wave component due to large momentum characters brought by the tensor force \cite{ikeda10}.  
This effect originates from the pseudo-scalar nature of the one-pion exchange.  
It is reported experimentally that a large fraction of $pn$ pairs is observed as compared with $pp$ or $nn$ pairs in light nuclei \cite{subedi08, simpson11}.  
This enhancement of $pn$ pairs is hard to reproduce theoretically in a simple shell model (mean-field picture) \cite{simpson11} except for the rigorous method as the one of GFMC \cite{schiavilla07}, 
which treats the tensor force explicitly. 
It is important to make efforts to study high momentum components caused by the tensor force in finite nuclei \cite{tanihata10}.

There are two important developments to perform nuclear structure calculations in heavy nuclei by including the necessary dynamics induced by the bare $NN$ interaction. 
One development is to find out that the strong tensor force is of intermediate range and 
we are able to express the tensor correlation in a reasonable shell model space~\cite{myo07,myo09,myo11}.  
We name this method as Tensor Optimized Shell Model (TOSM).  
The other is the Unitary Correlation Operator Method (UCOM) to treat the short-range correlation caused by the short-range repulsion~\cite{feldmeier98, neff03, roth10}. 
We shall combine these two methods, TOSM and UCOM, to describe nuclei using a bare $NN$ interaction.  
In the TOSM part, the wave function is constructed in terms of the shell model basis states with full optimization of two particle-two hole ($2p2h$) states. 
There is no truncation of the particle states within the $2p2h$ space in TOSM, where the spatial shrinkage of the particle states is essential to obtain convergence of the tensor contribution 
involving high momentum components~\cite{toki02,sugimoto04,ogawa06}.
This treatment of the bare tensor force in TOSM corresponds to the one-pair approximation correlated by the tensor force~\cite{togashi07,ogawa06,horii11}.  
In a few-body framework, the validity of TOSM was confirmed by taking only the $D$-wave component connected with the $S$-wave state directly by the tensor force for the $s$-shell nuclei \cite{horii11}.  
Their calculation reproduces more than 90\% of the tensor contribution and this result justifies the description of the tensor correlation in TOSM.
The explicit inclusion of $2p2h$ states in the extended mean field model for heavy nuclei has been formulated by Ogawa {\it et al.} as an extended Brueckner Hartree-Fock theory \cite{ogawa11}.

So far, we obtained various successful results of TOSM for the investigation of the tensor correlations in He isotopes.
In $^4$He, we showed the selectivity of $(p_{1/2})^2(s_{1/2})^{-2}$ configuration in the $2p2h$ space with the $pn$ pair induced by the tensor force and 
this correlation was recognized as the deuteron-like state \cite{myo09}.  
This specific $2p2h$ excitation plays a decisive role to reproduce the spectra of neutron-rich He isotopes \cite{myo11,myo05,myo06}.
It is found that the $p_{3/2}$ occupation of extra neutrons increases the tensor correlation of nuclei from $^5$He to $^8$He,
while the $p_{1/2}$ occupation of extra neutrons decrease the tensor correlation of those nuclei
due to the Pauli blocking between the specific $2p2h$ excitations by the tensor force in $^4$He and the motions of the extra $p_{1/2}$-neutrons. 
This configuration dependence of the tensor correlation produces the right amount of splitting energy between the $p_{3/2}$- and $p_{1/2}$-dominant states in the He isotopes.

In this paper, we use the TOSM+UCOM for the study of the Li isotopes from $^5$Li to $^9$Li, and
discuss their structures focusing on the role of the tensor force on the excitation energies and configurations of those nuclei. 
Before the present analysis, for neutron-rich $^{10}$Li and $^{11}$Li, we have performed the coupled two- and three-body cluster model analyses 
assuming the $^9$Li core described using the simplified TOSM wave function, respectively \cite{ikeda10,myo07_11,myo08}.
The cluster model for two nuclei precisely describes the neutron wave functions in a loosely bound state around $^9$Li, such as a neutron halo structure in $^{11}$Li.
Similar to the case of the He isotopes, we have shown that $^9$Li has the amount of the selected excitation from the $0s$ orbit to the $0p_{1/2}$ orbit by the tensor force,
and the $p_{1/2}$ occupation of last one and two neutrons in $^{10}$Li and $^{11}$Li are blocked by this tensor correlation in $^9$Li. 
As a result, the $1s$ occupation of last neutrons is relatively enhanced in two nuclei, and the virtual $s$-state in $^{10}$Li, and the large $s$-wave mixing and a neutron halo formation in $^{11}$Li are naturally explained.
On the basis of these successful results for two nuclei, in the present study, we perform a full microscopic analysis by using the TOSM+UCOM approach using the $NN$ interaction for the Li isotopes.
We investigate the systematic role of the tensor force in $^{5-9}$Li. 
It is interesting to see the effect of tensor force on the structures of $^{5,6}$Li \cite{arai95,kikuchi11}.
In $^6$Li, the isospin $T=0$ and $T=1$ states coexist in the low excitation energy region.
The influence of the tensor force on two kinds of states in $^6$Li are important to understand the isospin dependence of $^6$Li in relation with the tensor correlation.
In heavier Li isotopes, the contribution of tensor force is investigated in the ground and excited states. 
We also see the high momentum behavior of the kinetic energy in each state in relation with the tensor correlation. 
The present subject of Li isotopes becomes a foundation of the nuclear structure study using TOSM+UCOM and also of the previous study of $^{11}$Li and $^{10}$Li structures \cite{myo07_11,myo08}.

In Sec.~\ref{sec:model}, we explain the method of the TOSM+UCOM approach.  
In Sec.~\ref{sec:result}, we show the level structures of Li isotopes and discuss their characteristics in relation with the tensor force.  
A summary is given in Sec.~\ref{sec:summary}.

\section{Theoretical Framework}\label{sec:model}

\subsection{Tensor-optimized shell model (TOSM)}

We explain the tensor optimized shell model (TOSM) for open shell nuclei. We begin with writing a many-body Hamiltonian for a $A$ body system as,
\begin{eqnarray}
    H
&=& \sum_i^{A} T_i - T_{\rm c.m.} + \sum_{i<j}^{A} V_{ij} , 
    \label{eq:Ham}
    \\
    V_{ij}
&=& v_{ij}^C + v_{ij}^{T} + v_{ij}^{LS} + v_{ij}^{Clmb} .
\end{eqnarray}
Here, $T_i$ is the kinetic energy of each nucleon with $T_{\rm c.m.}$ being the center of mass kinetic energy.  
We take a bare interaction $V_{ij}$ such as AV8$^\prime$ \cite{pudliner97} consisting of central $v^C_{ij}$, tensor $v^T_{ij}$ and spin-orbit $v^{LS}_{ij}$ terms and $v_{ij}^{Clmb}$ the Coulomb term.  
We obtain the many-body wave function $\Psi$ by solving the Schr\"odinger equation $H \Psi=E \Psi$.  In our previous works for $^{4}$He and a few body systems, we found that the tensor force can be described by taking $2p2h$ excitations with large momentum components in the shell-model framework \cite{myo09,horii11}.

We prepare first a standard shell-model state for an open shell Li nucleus with $A$ nucleons in order to introduce the TOSM for open shell nuclei. The standard shell-model state $\Psi_{S}$ is written as
\beq
\Psi_{S}=\sum_{k_{S}} A_{k_{S}}|(0s)^{4}(0p)^{A-4};k_{S}\ket~.
\eeq
Here, the $p$ shell is the valence shell and the index $k_S$ is used to distinguish various shell-model components.
The spirit of the TOSM is that the tensor force works strongly for two nucleons in the standard shell-model state and excite two nucleons to various two particle states with high momentum components.  Hence, we limit configurations up to two particle-two hole excitations from the standard shell-model state.

We extend therefore first the standard shell-model state by allowing two particles in the $s$ shell to excite into the $p$ shell.  We name these extended shell-model states as $\kets{0p0h;k_{0}}$ states, since no particle is excited into shell-model orbits outside of the $sp$ shells.   We write the extended shell-model states as
\beq
\kets{0p0h;k_0}=|(0s)^{n_{s}}(0p)^{n_{p}};k_{0}\ket
\eeq
with the constraints $n_{s}+n_{p}=A$ and $2 \le n_{s} \le 4$.  $k_{0}$ distinguishes various $sp$ shell configurations.   We take now all the necessary high momentum components brought by the strong tensor force. These components are included by exciting two nucleons from the $sp$ shells to higher shells and therefore limit these configurations to $2p2h$ states from the standard shell-model state $\Psi_{S}$.  Hence, we have
\beq
\kets{2p2h;k_{2}}=|(0s)^{n_{s}}(0p)^{n_{p}}(higher)^{2};k_{2}\ket
\eeq
with the constraint $n_{s}+n_{p}=A-2$ and $2 \le n_{s} \le 4$.  We introduce quantum numbers $k_{2}$ to distinguish various $2p2h$ states, which amount to a large number of configurations.
Here, ``{\it higher}'' indicates higher shells outside of the $sp$ shells, which are treated as particle states.  Hence, we use the notation of $2p2h$ states to specify that two particles are in higher shells outside of the $sp$ shells.
We allow also $1p1h$ excitations for shell model consistency.
\beq
\kets{1p1h;k_{1}}=|(0s)^{n_{s}}(0p)^{n_{p}}(higher)^{1};k_{1}\ket
\eeq
with the constraint $n_{s}+n_{p}=A-1$ and $2 \le n_{s} \le 4$.  These $1p1h$ states are able to include high momentum components and improve the $0p0h$ wave functions in the presence of the strong tensor correlation.  Altogether, we write the TOSM wave function $\Psi$ for open shell nuclei as 
\begin{eqnarray}
\Psi&=& \sum_{k_0} A_{k_0} \kets{0p0h;k_0} + \sum_{k_1} A_{k_1} \kets{1p1h;k_1}
\nonumber\\
&+& \sum_{k_2} A_{k_2} \kets{2p2h;k_2}.
      \label{eq:config}
\end{eqnarray}
Here, all the amplitudes $\{A_{k_0},A_{k_1},A_{k_2}\}$ are variational coefficients to be fixed by the energy minimization.

We write now the details of the radial wave functions.  The $0p0h$ states are shell model states and expressed in terms of harmonic oscillator wave functions.  Hence, the $0s$ and $0p$ shell-model wave functions are used to express the radial wave functions, whose length parameters are taken independently as variational parameters.
The $1p1h$ and $2p2h$ states involve particle states with high momentum components.  The hole states correspond to the shell-model states in the $sp$ shells.  The particle wave functions have to contain high momentum components to express the specific characters of the tensor force with all the possible angular momenta until the total energy converges.

For particle states, we employ the Gaussian wave functions to express single-particle states in higher orbits in order to describe high momentum properties due to the tensor force~\cite{hiyama03, aoyama06}.  We prefer the Gaussian wave functions over the shell-model states in order to effectively include the necessary high momentum components \cite{myo09}.
When we superpose a sufficient number of Gaussian wave functions, the radial components of the particle states can be fully expressed.
Gaussian basis states should be orthogonalized to the hole states and among themselves.
This condition is imposed by using the Gram-Schmidt orthonormalization. 
In order to use the non-orthogonal Gaussian basis functions in the shell model framework, we construct the following orthonormalized single-particle basis function $\psi^n_{\alpha}$ using a linear combination of Gaussian bases $\{\phi_\alpha\}$ with length parameter $b_{\alpha,\nu}$.
\begin{eqnarray}
        \psi^n_{\alpha}(\vc{r})
&=&     \sum_{\nu=1}^{N_\alpha} d^n_{\alpha,\nu}\ \phi_{\alpha}(\vc{r},b_{\alpha,\nu}),
        \label{eq:Gauss1}
        \\
        \bra \psi^n_{\alpha} | \psi^{n'}_{\alpha'}\ket
&=&     \delta_{n,n'}\ \delta_{\alpha,\alpha'},
        \label{eq:Gauss3}
	\\
{\rm for}~~n~&=&~1,\cdots,N_\alpha,       
        \nonumber
\end{eqnarray}
where $N_\alpha$ is a number of basis functions for the orbit $\alpha$, and $\nu$ is an index to distinguish the bases with Gaussian length of $b_{\alpha,\nu}$.
The explicit form of the Gaussian basis function is written as
\begin{eqnarray}
        \phi_{\alpha}(\vc{r},b_{\alpha,\nu})
&=&     N_l(b_{\alpha,\nu}) r^l e^{-(r/b_{\alpha,\nu})^2/2} [Y_{l}(\hat{\vc{r}}),\chi^\sigma_{1/2}]_j ,
        \label{eq:Gauss2}
        \\
        N_l(b_{\alpha,\nu})
&=&     \left[  \frac{2\ b_{\alpha,\nu}^{-(2l+3)} }{ \Gamma(l+3/2)}\right]^{\frac12},
\end{eqnarray}
where $l$ and $j$ are the orbital and total angular momenta of the basis states, respectively. 
The weight coefficients $\{d^n_{\alpha,\nu}\}$ are determined to satisfy the overlap condition in Eq.~(\ref{eq:Gauss3}). 
This is done by using the Gram-Schmidt orthonormalization. 
Following this procedure, we obtain the new single-particle basis states $\{\psi^n_{\alpha}\}$ in Eq.~(\ref{eq:Gauss1}) used in TOSM. 
The particle states in $1p1h$ and $2p2h$ states are prepared to specify the basis wave functions, whose amplitudes are determined by the variational principle.  

We note that we may use another method to obtain $\{\psi^n_{\alpha}\}$ by solving the eigenvalue problem of the norm matrix of the Gaussian basis set in Eq.~(\ref{eq:Gauss2}) with the dimension $N_\alpha$.
This method gives different coefficients $\{d^n_{\alpha,\nu}\}$ from those of the Gram-Schmidt method. However, these two methods of making the orthonormalized single-particle basis states provide equivalent variational solutions for the total TOSM wave function $\Psi$ in Eq.~(\ref{eq:config}), because we start from the same Gaussian basis functions in Eq.~(\ref{eq:Gauss2}) with the same number $N_\alpha$ of Gaussian basis states.
Therefore, the TOSM solution does not depend on the construction of the orthonormal basis states when we start from the Gaussian basis functions.

We construct Gaussian basis functions of particle states to be orthogonal to the occupied states.
For the $1s$ orbit in the particle states, we prepare an extended $1s$ basis function \cite{myo07,myo09}, which is orthogonal to the $0s$ state and possesses a length parameter $b_{1s,\nu}$ that can differ from $b_{0s}$ of the $0s$ state.
In the extended $1s$ basis functions, the polynomial part is changed from the usual $1s$ basis states to satisfy the conditions of the normalization and the orthogonality to the $0s$ state \cite{myo07}. 
For the $1p$ orbits, we take the same method as used for the $1s$ case. 
In the numerical calculation, we prepare 10 Gaussian basis functions at most with various range parameters to get a convergence of the energy and Hamiltonian components.

We note here that when we write probabilities and occupation numbers of each orbit in various states in the numerical sections, 
those numbers are given by the summation of all the orthogonal orbits with the same spin having different radial behaviors due to the fact these wave functions are constructed by the orthonormalization.
For hole states, we do not have to sum up their numbers because the hole states are described by the single harmonic oscillator basis states.

We take care of the center-of-mass excitations by using the well-tested method of introducing a Hamiltonian of center-of-mass motion in the many-body Hamiltonian known as the Lawson method~\cite{lawson}. 
In the present study, we take the value of $\hbar \omega$ for the center of mass motion as the averaged one used in the $0s$ and $0p$ orbits in the $0p0h$ states with the weight of the occupation numbers in each orbit \cite{myo11}.  
Adding this center of mass Hamiltonian as the Lagrange multiplier to the original Hamiltonian in Eq.~(\ref{eq:Ham}), we can effectively project out only the lowest harmonic oscilattor state for the center-of-mass motion.  

The variation of the energy expectation value with respect to the total wave function $\Psi$ in Eq.~(\ref{eq:config}) is given by
\begin{eqnarray}
\delta\frac{\bra\Psi|H|\Psi\ket}{\bra\Psi|\Psi\ket}&=&0\ ,
\end{eqnarray}
which leads to the following equations:
\begin{eqnarray}
    \frac{\del \bra\Psi| H - E |\Psi \ket} {\del b_{\alpha,\nu}}
&=& 0\ ,\quad
   \label{eq:vari1}
    \\
    \frac{\del \bra\Psi| H - E |\Psi \ket} {\del A_{k_i}}
&=&  0\qquad \mbox{for}~i=0,1,2 .
   \label{eq:vari2}
\end{eqnarray}
The total energy is represented by $E$.
We solve two variational equations in Eqs. (\ref{eq:vari1}) and (\ref{eq:vari2}) in the following steps.  
First, fixing the length parameters $b_{\alpha,\nu}$ and the partial waves of the basis states up to $L_{\rm max}$, 
we solve the linear equation for $\{A_{k_i}\}$ as an eigenvalue problem for $H$. 
We thereby obtain the eigenvalue $E$, which is functions of $\{b_{\alpha,\nu}\}$ and $L_{\rm max}$. 
Next, we try to adopt various sets of the length parameters $\{b_{\alpha,\nu}\}$ and increase $L_{\rm max}$ in order to find a better solution which minimizes the total energy $E$. 
In TOSM, we can describe the spatial shrinkage of particle states with an appropriate radial form in the individual configuration, 
which is important to describe the tensor correlation \cite{myo07}, as seen in the deuteron.

\subsection{Unitary Correlation Operator Method (UCOM)}

We briefly explain UCOM for the short-range central correlation \cite{feldmeier98,neff03,roth10}, 
in which the following unitary operator $C$ is introduced
\begin{eqnarray}
C     &=&\exp(-i\sum_{i<j} g_{ij})~.
\label{eq:ucom}
\end{eqnarray}
We express the correlated wave function $\Psi$ in terms of a simple wave function $\Phi$ as $\Psi=C\Phi$. 
The transformed Schr\"odinger equation becomes $\hat H \Phi=E\Phi$ where the transformed Hamiltonian is given as $\hat H=C^\dagger H C$.   
The operator $C$ is in principle a many-body operator. 
In case of the short-range correlation, we are able to truncate the modified operators at the level of two-body operators~\cite{feldmeier98}.

Two-body Hermite operator $g$ in Eq.~(\ref{eq:ucom}) is defined as
\begin{eqnarray}
g &=& \frac12 \left\{ p_r s(r)+s(r)p_r\right\} ~,
\label{eq:ucom_g}
\end{eqnarray}
where the operator $p_r$ is the radial component of the relative momentum and is conjugate to the relative coordinate $r$. 
The function $s(r)$ expresses the amount of the shift of the relative wave function at the relative coordinate $r$ for every nucleon pair in nuclei.  
We use the TOSM basis states to describe $\Phi$.

In UCOM, the function $s(r)$ is determined variationally to minimize the total energy of the system. 
We parametrize $s(r)$ in the same manner as proposed by Feldmeier and Neff \cite{feldmeier98,neff03} for four channels of spin-isospin pair.
The detailed forms of $s(r)$ and their parametrization are explained in the previous paper \cite{myo11}.
In the present analysis, we use these $s(r)$ functions commonly for every states of the Li isotopes.  
To simplify the numerical calculation, we adopt the ordinary UCOM for the central correlation part instead of the $S$-UCOM in this analysis.

\section{Numerical Results}\label{sec:result}

\subsection{$^4$He}

We explain first the results of $^4$He calculated by using TOSM+UCOM.  We then discuss how the structures of $^{5-9}$Li behave by comparing with the properties of $^4$He.
We show results of $^4$He using the AV8$^\prime$ interaction, which consists of central, $LS$ and tensor terms.  The AV8$^{\prime}$ interaction is used in the calculation given by Kamada {\it et al}, where the Coulomb term is ignored \cite{kamada01}.  
The total energy of $^4$He is obtained as $-22.30$ MeV in TOSM+UCOM.
The convergence of the solutions of $^4$He has already been confirmed in the previous works~\cite{myo11,myo09}.
The energy components of $^4$He are discussed in the previous paper~\cite{myo11}

\begin{table}[t]
\caption{Probabilities of various configurations in $^4$He with TOSM+UCOM using the AV8$^\prime$ interaction, where the two subscripts $00$ and $10$ are the spin-isospin quantum numbers.}
\begin{tabular}{c|r}
\noalign{\hrule height 0.5pt}  
$(0s)_{00}^4$                                  & 84.14  \\
$(0s)_{10}^{-2}(0p_{1/2})_{10}^2$              &  2.32  \\
$(0s)_{10}^{-2}[(1s_{1/2})(0d_{3/2})]_{10}$    &  2.20  \\
$(0s)_{10}^{-2}[(0p_{3/2})(0f_{5/2})]_{10}$    &  1.82  \\
$(0s)_{10}^{-2}[(0p_{1/2})(0p_{3/2})]_{10}$    &  1.21  \\
\noalign{\hrule height 0.5pt}
\mbox{remaining part}                          &  8.31  \\
\noalign{\hrule height 0.5pt}
\end{tabular}
\label{conf4}
\end{table}

The dominant configurations of $^4$He in TOSM are listed in Table \ref{conf4} with their probabilities. 
The specific $2p2h$ states such as $(0s)_{10}^{-2}(0p_{1/2})_{10}^2$ show large probabilities and are essential to produce the tensor correlation in $^4$He.
The selectivity of those $2p2h$ configurations can be understood from the coupling by the tensor operator which changes both the orbital angular momentum and spin 
by two with the opposite direction of their $z$-components \cite{myo07, toki02}.  
In addition, the $2p2h$ state with $(0s)_{10}^{-2}[(1s_{1/2})(0d_{3/2})]_{10}$ is also favored by the tensor interaction.
It is found that the dominant $2p2h$ states commonly correspond to the excitations of a $pn$ pair.
Those features of the $2p2h$ excitations play an important role to determine the structures of $^{5-9}$Li as will be discussed later. 

For comparison, we show the results with the Minnesota (MN) $NN$ interaction, which does not have the tensor force.
The MN interaction consists of the central and $LS$ parts with $u$ parameter being 0.95 \cite{reichstein70, tang78}. We use set III of the MN interaction for the $LS$ force.
TOSM has been confirmed to reproduce the binding energy and the radius of $^4$He obtained by few-body stocastic variational method\cite{varga95,suzuki08,myo11}.
It is obtained that the $(0s)^4$ configuration is by far dominant in the $^4$He wave function by 96.6\%.  
Other $2p2h$ configurations have the probability less than a percent. 
This comparison makes the role of the tensor force clear on the energy spectra of the $p$-shell nuclei for the case of the AV8$^\prime$ interaction.

We also list the occupation numbers of nucleon in $^4$He using AV8$^\prime$ in Table \ref{occ4}.
It is shown that the $p_{1/2}$ orbit has the largest contribution among the particle states according to the large $2p2h$ mixing including the $p_{1/2}$ component.  
In the MN case, it is found that the component of the $0s$ orbit  is larger than the AV8$^\prime$ case and the enhancement of the $p_{1/2}$ orbit is not obtained.
These results mean that the tensor force brings the specific excitations from the $s$-shell to the $p$- and $sd$-shells in $^4$He.

\begin{table}[t]  
\caption{Occupation numbers in each orbit of $^4$He using AV8$^\prime$ and Minnesota (MN) interactions.}
\begin{tabular}{c|cccccccc}
\noalign{\hrule height 0.5pt}
$^4$He$(J^\pi)$  &~$0s_{1/2}$~&~$p_{1/2}$~&~$p_{3/2}$~&~$1s_{1/2}$~&~$d_{3/2}$~&~$d_{5/2}$~\\ 
\noalign{\hrule height 0.5pt}                                                 
AV8$^\prime$     & ~3.77~     &~ 0.06 ~   & ~ 0.04 ~  &~ 0.03 ~    &~ 0.04  ~  &~ 0.01~    \\ 
MN               & ~3.94~     &~ 0.01 ~   & ~ 0.03 ~  &~ 0.01 ~    &~ 0.004 ~  &~ 0.006~   \\ 
\noalign{\hrule height 0.5pt}
\end{tabular}
\label{occ4}
\end{table}

\subsection{Energy spectra of the Li isotopes}

\begin{figure}[t]
\centering
\includegraphics[width=8.5cm,clip]{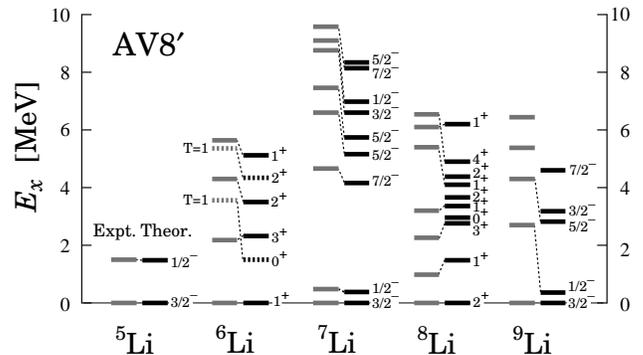}
\caption{Excitation energies of Li isotopes with TOSM+UCOM using AV8$^\prime$.} 
\label{fig:AV_Li}
\end{figure}

\begin{figure}[t]
\centering
\includegraphics[width=8.0cm,clip]{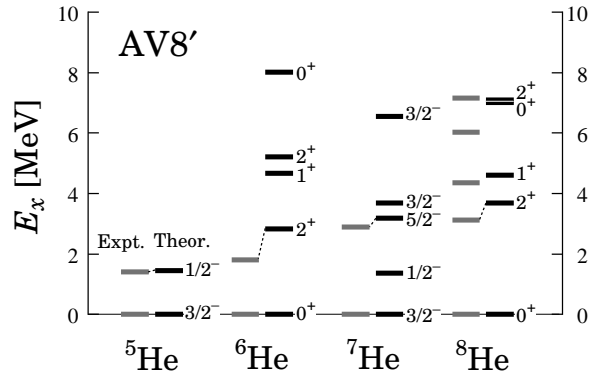}
\caption{Excitation energies of He isotopes using AV8$^\prime$ with TOSM+UCOM.} 
\label{fig:AV_He}
\end{figure}

\begin{figure}[t]
\centering
\includegraphics[width=8.5cm,clip]{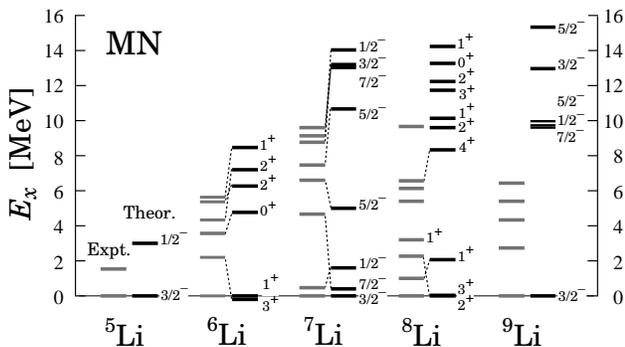}
\caption{Excitation energies of the Li isotopes using the Minnesota interaction (MN) with TOSM.}
\label{fig:MN}
\end{figure}

\begin{table}[t]
\caption{Matter radii of Li isotopes in comparison with the experiments; a\cite{tanihata88}, b\cite{dobrovolsky06}. Units are in fm.}
\label{radius}
\centering
\begin{tabular}{r|p{1.5cm} p{4.0cm}}
\hline
          & TOSM  &  Experiment        \\
\hline
$^6$Li    &~~2.22 &  2.35(3)$^{\rm a}$~~~~2.44(7)$^{\rm b}$ \\
$^7$Li    &~~2.32 &  2.35(3)$^{\rm a}$                      \\
$^8$Li    &~~2.42 &  2.38(2)$^{\rm a}$~~~~2.50(6)$^{\rm b}$ \\
$^9$Li    &~~2.46 &  2.32(2)$^{\rm a}$~~~~2.44(6)$^{\rm b}$ \\
\hline
\end{tabular}
\end{table}

\begin{figure}[t]
\centering
\includegraphics[width=8.5cm,clip]{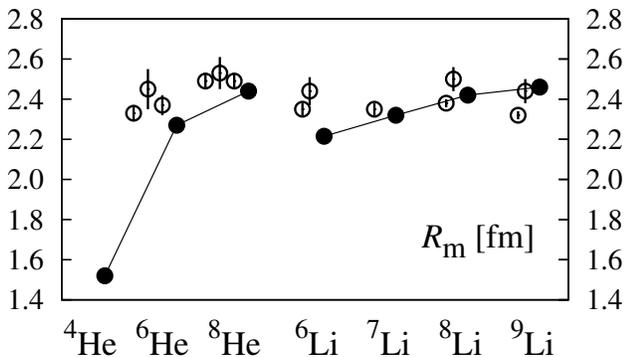}
\caption{Matter radius ($R_m$) of the He and Li isotopes in TOSM+UCOM with solid circles. Open circles are the experiments;
\cite{tanihata92,alkazov97,kiselev05} for He isotopes and \cite{tanihata88,dobrovolsky06} for Li isotopes.}
\label{fig:radius}
\end{figure}

We show the calculated results of the Li isotopes using TOSM+UCOM with the AV8$^\prime$ interaction, where $L_{\rm max} $ is taken as 10 to get sufficient convergence. 
We show the excitation spectra of $^{5-9}$Li in Fig. \ref{fig:AV_Li}.   We see quite a good correspondence to the experimental spectra. 
Similar to this result, we obtain good results for the excitation energies in the He isotopes as shown in Fig. \ref{fig:AV_He}, the details of which were discussed in Ref. \cite{myo11}.
In Fig. \ref{fig:AV_Li}, the resulting level spacing of the Li isotopes in TOSM+UCOM is good, but slightly more compact than the  experimental spectra.
For example, in $^9$Li, the small energy difference between the lowest $3/2^-$ and $1/2^-$ states in TOSM+UCOM in comparison with the experiment.
These characteristics are commonly obtained in the GFMC calculation \cite{pieper02}.  We will discuss in detail all the level structures of the Li isotopes in the next sub-section.
The additional genuine three-body interaction can be one of the components to reproduce the experimental situation \cite{pieper01}.
As for total binding energies, our results underestimate them in He and Li isotopes and its amount becomes larger for neutron-rich side. 

\begin{figure}[t]
\centering
\includegraphics[width=8.5cm,clip]{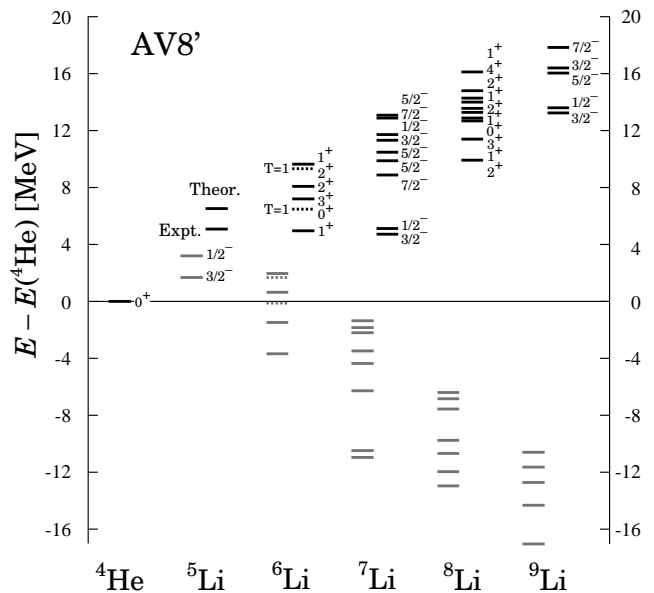}
\caption{Total energies of Li isotopes using AV8$^\prime$ measured from that of $^4$He.  The experimental energy spectra are shown in the left hand side of the corresponding theoretical spectra.} 
\label{total_ene}
\end{figure}

To see the interaction dependence, the energy spectra using the MN interaction are shown in Fig. \ref{fig:MN}. 
In the MN case, the excitation energies are split in two regions in comparison with the experiments.
One of the reasons is the too large splitting energy between $p_{3/2}$ and $p_{1/2}$ components mainly given by the $LS$ interaction. 
When we adjust the MN interaction such as changing the $u$ parameter and the strength of the $LS$ interaction, to fit the $LS$ splitting energy of $^5$Li,
it is still difficult to reproduce the whole trend of the energies of the ground and excited states of Li isotopes consistently.  A large difference between the AV8$^\prime$ and MN interactions is the tensor force.  The comparison of the energy spectra of the two interactions indicates the effect of the tensor force, which has a large impact on the excitation energy spectra of the He and Li isotopes.

We show total energies of the Li isotopes as shown in Fig.~\ref{total_ene}.  The experimental total energies become lower than that of $^{4}$He as the neutron number increases.  On the other hand, the calculated results go up in energy with the neutron number.  Hence, we lack some attractions, which increases with the mass number.   This luck of mass dependent attraction was already seen in the case of the He isotopes.
In the previous analysis of the He isotopes \cite{myo11}, we have already discussed the reasons of underbinding, which can contribute to the bulk part of the binding energies; \\
1) higher configurations beyond the $2p2h$ states in TOSM, such as the $4p4h$ states to include the two-kinds of the $2p2h$ excitations with isospin $T=0$ and $T=1$ pairs simultaneously.\\
2) the genuine three-body interaction, and \\
3) the improvement of the correlation function form of $R_+(r)$ in UCOM.

We have estimated these three effects on the binding energies of He isotopes such as $^6$He and $^8$He
and found that we can obtain sufficient binding energies for He isotopes, which are comparable with the experimental values \cite{myo11}.
In the present study, from the results of energy spectra in Figs. \ref{fig:AV_Li} and \ref{fig:AV_He}, we focus on the discussion of the structure differences between individual energy levels of the Li isotopes.
We show also the matter radii of $^{6,7,8,9}$Li in TOSM+UCOM in Table \ref{radius}, which agree with experiments.
We include the results of He isotopes in addition to the Li case in Fig. \ref{fig:radius} and find the whole trend on the matter radii observed in the He and Li isotopes is very good.

\subsection{Energy components and configurations}

\subsubsection{$^5$Li}

We discuss in detail the structures of each energy level of the Li isotopes obtained by using the TOSM+UCOM framework with the AV8$^\prime$ $NN$ interaction.  In Fig. \ref{fig:AV_Li}, we show the energy spectrum of $^5$Li as compared with experiment, where the energy difference of the spin-orbit partner states is obtained as $\Delta E=1.7$ MeV. 
It is noted that in the present calculation, the continuum effect of the last unbound proton is not included because of the bound state approximation.
It is naively expected that the inclusion of the continuum state of a last proton can reduce the splitting energy of $^5$Li because of the spatial extension of the proton wave function in the asymptotic region.
There is also a discussion of the effect of the genuine three-body interaction on the splitting energy in the $^4$He+$n$ scattering \cite{nollett07}.

In Table \ref{ham5}, we compare various energy components in the $3/2^-$ and $1/2^-$ states of $^5$Li measured from those of $^4$He.  We discuss the double roles of the tensor force on these spin-orbit partner states.
From Table \ref{ham5}, it is found that the tensor energy of the $3/2^-$ state is attractive, while it is very weak for $1/2^-$. 
The larger contribution of the tensor force in $3/2^-$ results in the enhancement of the kinetic energy because of the involvement of high momentum components brought by the tensor force.  
For $1/2^-$, on the other hand, the energy gain from the tensor force is small and the enhancement of the kinetic energy is small. 
These differences are related to the large mixing of the $p_{1/2}$ component in $^4$He as shown in Table \ref{conf4}.
In $^5$Li, when the last proton occupies the $p_{3/2}$ orbit, the tensor correlation in $^4$He is not much reduced.  
On top the $p_{3/2}$ state has an additional tensor contribution because of the tensor correlation between the $p_{3/2}$ proton and the neutrons in the $s$-shell.
In case of the $p_{1/2}$ occupation, this proton blocks the spatially compact component of the $p_{1/2}$ proton in the $^4$He part because of the small degeneracy of the $p_{1/2}$ orbit. 
This dynamics produces the Pauli-blocking and reduces the total binding energy of $^5$Li.  
As a result, the last proton occupied in the $p_{1/2}$ orbit should be orthogonal to the excited $p_{1/2}$ orbit in $^4$He and the tensor force does not gain the energy in $^5$Li($1/2^-$). 
Those different couplings of the tensor force are essential to explain the energy components in two states of $^5$Li and results in the $LS$ splitting energy as a net value 
\cite{terasawa59,arima60,nagata59,myo11}.

\begin{table}[t]
\caption{Various energy components in $^5$Li measured from those of the $^4$He ground state. Units are in MeV.}
\begin{tabular}{ccccccc}
\noalign{\hrule height 0.5pt}
$J^\pi$    &  Kinetic  &   Central &   Tensor  &     $LS$   \\
\noalign{\hrule height 0.5pt}
3/2$^-$    &  $ 10.96$ & $  -2.00$ & $  -3.07$ & $  -1.74$  \\
1/2$^-$    &  $ ~5.90$ & $ ~~0.30$ & $  -0.41$ & $ ~~0.03$  \\
\noalign{\hrule height 0.5pt}
\end{tabular}
\label{ham5}
\end{table}

\subsubsection{$^6$Li}
\begin{table}[t]
\caption{Various energy components in $^6$Li measured from those of the $^4$He ground state. Units are in MeV.}
\begin{tabular}{ccccccc}
\noalign{\hrule height 0.5pt}
$J^\pi$($T$) & Kinetic   & Central  & Tensor   & $LS$     \\
\noalign{\hrule height 0.5pt}
$1^+_1$ (0)  & $28.29$   & $-9.76$  & $-11.81$ & $-2.28$  \\
$3^+$   (0)  & $27.57$   & $-8.75$  & $-8.75$  & $-3.35$  \\
$2^+$   (0)  & $25.90$   & $-8.08$  & $-8.69$  & $-1.29$  \\
$1^+_2$ (0)  & $23.96$   & $-6.63$  & $-6.44$  & $-1.32$  \\
\noalign{\hrule height 0.5pt}
$0^+$ (1)    & $25.72$   & $-13.35$ & $-4.12$  & $-2.29$  \\
$2^+$ (1)    & $30.50$   & $-12.00$ & $-6.64$  & $-3.07$  \\
\noalign{\hrule height 0.5pt}
\end{tabular}
\label{ham6}
\end{table}

In Fig. \ref{fig:AV_Li}, we show the energy spectrum of $^6$Li as compared with experiment.  We see the ground state has the spin-parity of $J^{\pi}=1^{+}$, while states with $T=1$ are closer to the $1^{+}$ state as compared to experiment.  In order to understand the reason of the calculated results, we show energy components in Table \ref{ham6} measured from the energy components in $^4$He.
It is found that the $1^+$ ground state mostly exhausts the tensor energy.
This is because of the deuteron-like correlations of the valence $pn$-pair in addition to that in the $^4$He part.
Corresponding to this fact, the kinetic energy of the $1^+_1$ state shows the largest value among the isospin $T=0$ state of $^6$Li.
The $2^+$ and $3^+$ states with $T=0$ almost show the similar energies for the central and tensor energy components.
For the two isospin $T=1$ states, corresponding to the isobaric analog states of $^6$He, the tensor energies are mostly much smaller than the $T=0$ states of $^6$Li.
This is naturally understood from the viewpoint of the isospin dependence of the tensor force.
In the $T=1$ states of $^6$Li, the $^4$He part contains the $pn$ pair excitation with $T=0$ from the $0s$-shell by the tensor force,
but the outer $pn$ pair in the $0p$-shell with a $T=1$ state shows a small amount of the excitation due to the weak tensor force in the $T=1$ channel.
The $2^+$ state with $T=1$ gains the larger tensor energy than those of $0^+$ state and show the larger kinetic energy.

For energy levels of $^6$Li in comparison with experiment, 
the TOSM results give a small energy difference between $T=0$ and $T=1$ states as shown in Fig. \ref{fig:AV_Li}.
This can be related with the clustering effect of $\alpha$+$d$ structure of $^6$Li with $T=0$.
The TOSM wave function is expanded in terms of the shell model basis states and hence, is naively difficult to describe the tail component of the spatially extended clustering states, such as the 3$\alpha$ state in the excited states of $^{12}$C \cite{ikeda80}.
Similarly, $^6$Li with the $T=0$ states is considered to need the clustering component of $\alpha$+$d$ in addition to the shell model like one \cite{arai95,kikuchi11}.
The lacking of the cluster components loses some amount of the binding energy of $^6$Li in TOSM.
On the other hand the $T=1$ states do not corresponds to the $\alpha$+$d$ clustering state, so that 
TOSM can describe those $T=1$ states, similar to the $^6$He case.
This structure difference between $T=0$ and $T=1$ states can explain the small energy difference between those states of $^6$Li in TOSM.
When we increase the strength of tensor force artificially by about 5\% in order to see the effect of tensor force on the $T=0$ and $T=1$ states in $^6$Li,
the energy gains of the $1^+$($T=0$) state is 3.1 MeV, which is larger than 2.7 MeV of the $0^+$($T=1$) state. 
This fact implies that the $T=0$ states in $^6$Li contains the tensor correlation much more due to the presence of the outer $pn$ pair with the $T=0$ channel.
The full inclusion of the $\alpha$+$d$ component in addition to the TOSM basis states would be one of the ways 
to explain the experimental energy difference between the $T=0$ and $T=1$ states of $^6$Li.

We list the dominant configurations of $^6$Li in Table \ref{conf6}.
For comparison of the $1^+_1$ and $1^+_2$ states, it is found that the $1^+_1$ state does not have the large component of the $(0p_{1/2})^2$ configuration of a $pn$ pair,
while the $1^+_2$ state has this configuration by 23\%.
This difference in the configurations explains the tensor energies in two states shown in Table \ref{ham6}.
Owing to the Pauli blocking between the $pn$-pair in the $0p$-shell and another pair excited from the $0s$-shell,
the $1^+_2$ suffers the large blocking effect due to the presence of the $(0p_{1/2})^2$ configuration,
which reduces the tensor contribution in this state.
The same state-dependence of the tensor correlation is mentioned in $^5$Li.

\begin{table}[t]
\caption{Dominant configurations of $^6$Li($J^\pi$($T$)) with their squared amplitudes $(A^J_k)^2$ using AV8$^\prime$ interaction.}
\label{conf6}
\begin{ruledtabular}
\begin{tabular}{ll|lrc}
\multicolumn{2}{c|}{$1^+_1(0)$}     & \multicolumn{2}{c}{$1^+_2(0)$}          \\ \hline
$(0s)^4(0p_{1/2})(0p_{3/2})$ & 0.43 & $(0s)^4(0p_{3/2})^2$            & 0.30  \\
$(0s)^4(0p_{3/2})^2$         & 0.38 & $(0s)^4(0p_{1/2})(0p_{3/2})$    & 0.29  \\
                             &      & $(0s)^4(0p_{1/2})^2$            & 0.23  \\
\end{tabular}
\vspace*{0.2cm}
\begin{tabular}{ll|lrc}
\multicolumn{2}{c|}{$2^+(0)$}             & \multicolumn{2}{c}{$3^+(0)$}      \\ \hline
$(0s)^4(0p_{1/2})(0p_{3/2})$       & 0.82 & $(0s)^4(0p_{3/2})^2$      & 0.82  \\
\end{tabular}
\vspace*{0.2cm}
\begin{tabular}{ll|lrc}
\multicolumn{2}{c|}{$0^+(1)$}      & \multicolumn{2}{c}{$2^+(1)$}         \\ \hline
$(0s)^4(0p_{3/2})^2$      & 0.72   & $(0s)^4(0p_{3/2})^2$          & 0.74 \\
$(0s)^4(0p_{1/2})^2$      & 0.11   & $(0s)^4(0p_{1/2})(0p_{3/2})$  & 0.11 \\ 
\end{tabular}
\end{ruledtabular}
\end{table}

\begin{table}[t]
\caption{Dominant configurations of $^6$Li($J^\pi$($T$)) using Minnesota interaction.}
\label{conf6_MN}
\begin{ruledtabular}
\begin{tabular}{ll|lrc}
\multicolumn{2}{c|}{$1^+_1(0)$}     & \multicolumn{2}{c}{$1^+_2(0)$}          \\ \hline
$(0s)^4(0p_{3/2})^2$         & 0.56 & $(0s)^4(0p_{1/2})(0p_{3/2})$    & 0.43  \\
$(0s)^4(0p_{1/2})(0p_{3/2})$ & 0.29 & $(0s)^4(0p_{3/2})^2$            & 0.19  \\
                             &      & $(0s)^4(0p_{1/2})^2$            & 0.11  \\
\end{tabular}
\vspace*{0.2cm}
\begin{tabular}{ll|lrc}
\multicolumn{2}{c|}{$2^+(0)$}             & \multicolumn{2}{c}{$3^+(0)$}      \\ \hline
$(0s)^4(0p_{1/2})(0p_{3/2})$       & 0.83 & $(0s)^4(0p_{3/2})^2$      & 0.88  \\
\end{tabular}
\vspace*{0.2cm}
\begin{tabular}{ll|lrc}
\multicolumn{2}{c|}{$0^+(1)$}      & \multicolumn{2}{c}{$2^+(1)$}         \\ \hline
$(0s)^4(0p_{3/2})^2$      & 0.82   & $(0s)^4(0p_{3/2})^2$          & 0.77 \\
$(0s)^4(0p_{1/2})^2$      & 0.04   & $(0s)^4(0p_{1/2})(0p_{3/2})$  & 0.09 \\ 
\end{tabular}
\end{ruledtabular}
\end{table}

\begin{table}[t]  
\caption{Occupation numbers in each orbit of $^6$Li using AV8$^\prime$ interaction.}
\begin{tabular}{c|cccccccc}
\noalign{\hrule height 0.5pt}
$J^\pi,T$       &~$0s_{1/2}$~&~$0p_{1/2}$~&~$0p_{3/2}$~&~$1s_{1/2}$~&~$d_{3/2}$~&~$d_{5/2}$~ \\
\noalign{\hrule height 0.5pt}
$1^+_1,0$       & ~3.74~    &~ 0.53 ~   & ~ 1.42 ~  &~  0.04  ~ &~  0.05  ~ &~ 0.04~     \\ 
$1^+_2,0$       & ~3.74~    &~ 0.89 ~   & ~ 1.08 ~  &~  0.05  ~ &~  0.05  ~ &~ 0.04~     \\ 
$2^+,0$         & ~3.75~    &~ 0.97 ~   & ~ 0.98 ~  &~  0.04  ~ &~  0.05  ~ &~ 0.04~     \\ 
$3^+,0$         & ~3.74~    &~ 0.03 ~   & ~ 1.94 ~  &~  0.05  ~ &~  0.05  ~ &~ 0.03~     \\ 
\noalign{\hrule height 0.5pt}
$0^+,1$         & ~3.73~    &~ 0.28 ~   & ~ 1.70 ~  &~  0.04  ~ &~  0.05  ~ &~ 0.03~     \\ 
$2^+,1$         & ~3.74~    &~ 0.15 ~   & ~ 1.83 ~  &~  0.04  ~ &~  0.05  ~ &~ 0.03~     \\ 
\noalign{\hrule height 0.5pt}
\end{tabular}
\label{occ6}
\end{table}

\begin{table}[t]  
\caption{Occupation numbers in each orbit of $^6$Li using MN interaction.}
\begin{tabular}{c|cccccccc}
\noalign{\hrule height 0.5pt}
 $J^\pi,T$      &~$0s_{1/2}$~&~$0p_{1/2}$~&~$0p_{3/2}$~&~$1s_{1/2}$~&~$d_{3/2}$~&~$d_{5/2}$~ \\
\noalign{\hrule height 0.5pt}
$1^+_1,0$       & ~3.85~    &~ 0.37 ~   & ~ 1.60 ~  &~  0.03  ~ &~ 0.03~ &~  0.04~       \\ 
$1^+_2,0$       & ~3.83~    &~ 0.76 ~   & ~ 1.08 ~  &~  0.05  ~ &~ 0.04~ &~  0.06~       \\ 
$2^+,0$         & ~3.85~    &~ 0.94 ~   & ~ 0.99 ~  &~  0.03  ~ &~ 0.04~ &~  0.04~       \\ 
$3^+,0$         & ~3.89~    &~ 0.02 ~   & ~ 1.92 ~  &~  0.04  ~ &~ 0.01~ &~  0.04~       \\ 
\noalign{\hrule height 0.5pt}                                                     
$0^+,1$         & ~3.85~    &~ 0.12 ~   & ~ 1.85 ~  &~  0.03  ~ &~ 0.02~ &~  0.05~       \\ 
$2^+,1$         & ~3.85~    &~ 0.12 ~   & ~ 1.84 ~  &~  0.04  ~ &~ 0.02~ &~  0.04~       \\ 
\noalign{\hrule height 0.5pt}
\end{tabular}
\label{occ6_MN}
\end{table}

We also compare the dominant configurations of $^6$Li using AV8$^\prime$ and MN interactions.
The MN case is shown in Table \ref{conf6_MN}.
In the $1^+$ ground state, the difference in the results of two interactions is clearly seen;  
the $(0p_{1/2})(0p_{3/2})$ configuration in the $0p$-shell of $^6$Li is enhanced in the AV8$^\prime$ case rather than the MN one.
This configuration is a characteristic feature of the $LS$ coupling scheme.
We discuss the relation between this configuration and the tensor force.
The $(0p_{1/2})(0p_{3/2})$ configuration of $pn$ in $^6$Li involves the 89\% of the spin triplet ($S=1$) component, while the $(0p_{3/2})^2$ configuration
involves 44\%. The $S=1$ component in the wave function is important to activate the tensor force and is related to the deuteron-like component in the $pn$ pair.
It is found that the $(0p_{1/2})(0p_{3/2})$ configuration in $^6$Li is favored to increase the tensor energy in the wave function.
Contrastingly, in the MN case, there is no explicit tensor force and the $(0p_{3/2})^2$ configuration becomes dominant, 
which can be understood naively from the $jj$-coupling scheme.
It is experimentally interesting to observe the amount of the mixings of $p_{1/2}$ and $p_{3/2}$ orbits in $^6$Li.

\begin{figure}[t]
\centering
\includegraphics[width=6.5cm,clip]{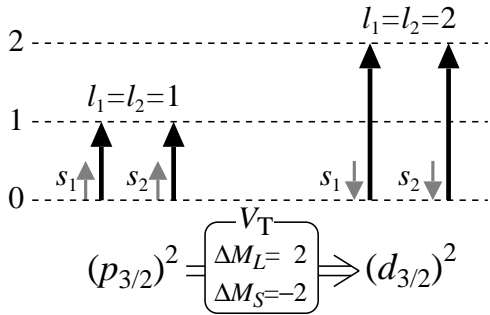}
\caption{Schematic figure to explain the selectivity of the tensor coupling of two nucleons from the $(p_{3/2})^2$ configuration to the $(d_{3/2})^2$ one.}
\label{fig:tensor}
\end{figure}

We list the occupation numbers of $^6$Li up to the $sd$-shell in Table \ref{occ6} using AV8$^\prime$. 
It is found that the occupation numbers in the $0s$-shell and the $sd$-shell show almost common values among the $^6$Li states,
and the $0p$-shell contributions depend on the states.
For comparison, we also list the MN case in Table \ref{occ6_MN}.
It is found that the $0s$ components in each state of $^6$Li using AV8$^\prime$ is larger than the MN case.
This is similar to the $^4$He results shown in Table \ref{occ4}.
Those results mean that the tensor force brings the excitation of the nucleon from the $0s$-orbit to higher orbits such as $p_{1/2}$ in $^6$Li.
This tendency can be commonly seen in every state of $^6$Li.
Among the $sd$-shell components of $^6$Li, it is found that the $d_{3/2}$ component is the largest one in most of the states using AV8$^\prime$. 
This aspect cannot be seen in the MN case shown in Table \ref{occ6_MN} , which shows the normal order of the $jj$ coupling scheme.
This difference comes from the tensor force, which selectively excites the nucleon in the $0p_{3/2}$ orbit in the hole state to the $d_{3/2}$ orbit in the particle state,
both of which has the same $j$ quantum number.
This reason is schematically explained in Fig.~\ref{fig:tensor}, in which the direction of the coupled orbital angular momentum and that of the coupled spin should be opposite in the tensor operator.
Considering the transition from the $(0p_{3/2})^2$ configuration by the tensor force, 
one of the favored components is naively estimated to be the $(0d_{3/2})^2$ one with the distribution of $l_1$=$l_2$=$2$,
where there are some fragment of the orbital angular momentum distribution of two nucleons in the particle states. 

The same selectivity of the tensor coupling can be seen in the case of $0s_{1/2}$-$0p_{1/2}$ combination in $^4$He.
These specific couplings can be seen from the view point of the pionic correlation in the particle-hole representation. 
The pion has the spin-parity of $0^-$.  Hence, hole configurations with positive or negative parity are favorably
excited to particle configurations with opposite parity with the same total spin due to the pion exchange.
The selectivity of the tensor force can be seen as the selectivity of the pion exchange \cite{toki02,ikeda10,myo05}. 
It is extremely interesting to observe experimentally the relatively large mixing of $d_{3/2}$ orbit among the $sd$-shell in $^6$Li.

\subsubsection{$^{7,8,9}$Li}

For $^7$Li, various energy components are shown in Table \ref{ham7} measured from the $^4$He components.
From the comparison of the components in each state, it is found that the $1/2^-_1$ and $3/2^-$ states show the similar structure having the large tensor contributions and also the larger kinetic energy than those of other states.

For comparison between the dominant configurations of $3/2^-_1$ and $3/2^-_2$ states from Table \ref{conf7},
the $3/2^-_1$ state does not have the $0p_{1/2}$ orbit dominantly and this property increases the tensor contribution via the excitation of $pn$-pair from the $0s$-shell.
On the other hand, $3/2^-_2$ state have $0p_{1/2}$ orbit in the most dominant configuration and this configuration blocks the excitation of $pn$-pair from the $0s$-shell
and cannot increase the tensor contribution in comparison with the $3/2^-_1$ state.

For comparison between the $1/2^-_1$ and $1/2^-_2$ states,
the $1/2^-_1$ state dominantly has the $nn$ pair of the $0p_{3/2}$ orbit with the $T=1$ component
and the $1/2^-_2$ state dominantly has the $pn$-pair of the $0p_{3/2}$ orbit with $T=0$.
The latter configuration in the $1/2^-_2$ state blocks the excitation of $pn$ pair from the $0s$-shell to $0p_{3/2}$ orbit with some amount, which reduces the tensor contribution.
These configurations in two $1/2^-$ states determine the different tensor contributions in two states.
From those analyses, it is found that the nucleons occupied in $0p$-shell in the lowest configuration play important role to explain the tensor contribution in each state of $^7$Li.

\begin{table}[t]
\caption{Various energy components in $^7$Li measured from those of the $^4$He ground state. Units are in MeV.}
\begin{tabular}{ccccccc}
\noalign{\hrule height 0.5pt}
$J^\pi$ & Kinetic& Central  & Tensor    & $LS$     \\
\noalign{\hrule height 0.5pt}
1/2$^-_1$    & 53.59 & $-28.75$ & $-16.45$ & $-2.56$ \\
1/2$^-_2$    & 46.85 & $-18.76$ & $-13.48$ & $-2.44$ \\
3/2$^-_1$    & 53.77 & $-29.46$ & $-15.62$ & $-3.24$ \\
3/2$^-_2$    & 46.95 & $-18.15$ & $-14.66$ & $-2.26$ \\
5/2$^-_1$    & 50.63 & $-22.95$ & $-14.74$ & $-2.40$ \\
5/2$^-_2$    & 49.35 & $-19.56$ & $-14.10$ & $-4.54$ \\
5/2$^-_3$    & 46.45 & $-17.58$ & $-13.02$ & $-2.64$ \\
7/2$^-_1$    & 52.74 & $-24.96$ & $-13.99$ & $-4.11$ \\
7/2$^-_2$    & 47.54 & $-18.02$ & $-12.82$ & $-3.76$ \\
\noalign{\hrule height 0.5pt}
\end{tabular}
\label{ham7}
\end{table}

\begin{table}[t]
\centering
\caption{Dominant configurations of $^7$Li($J^\pi$($T$)) with their squared amplitudes $(A^J_k)^2$.
Two subscripts are the spin-isospin quantum numbers.}
\label{conf7}
\begin{ruledtabular}
\begin{tabular}{ll|ll}
\multicolumn{2}{c|}{$1/2^-_1$}              & \multicolumn{2}{c}{$1/2^-_2$}               \\ \hline
$(0s)^4(0p_{3/2})^2_{01}(0p_{1/2})$  & 0.50 & $(0s)^4(0p_{3/2})^2_{10}(0p_{1/2})$ & 0.47  \\
$(0s)^4(0p_{3/2})^3$                 & 0.19 & $(0s)^4(0p_{3/2})^3$                & 0.17  \\
$(0s)^4(0p_{3/2})^2_{10}(0p_{1/2})$  & 0.07 & $(0s)^4(0p_{3/2})^2_{01}(0p_{1/2})$ & 0.09  \\ 
\end{tabular}
\vspace*{0.2cm}
\begin{tabular}{ll|ll}
\multicolumn{2}{c|}{$3/2^-_1$}              & \multicolumn{2}{c}{$3/2^-_2$}             \\ \hline
$(0s)^4(0p_{3/2})^3$                & 0.48  & $(0s)^4(0p_{3/2})^2_{21}(0p_{1/2})$ & 0.55 \\
$(0s)^4(0p_{3/2})(0p_{1/2})^2_{01}$ & 0.15  & $(0s)^4(0p_{3/2})^2_{10}(0p_{1/2})$ & 0.13 \\
$(0s)^4(0p_{3/2})^2_{21}(0p_{1/2})$ & 0.10  & $(0s)^4(0p_{3/2})(0p_{1/2})^2_{01}$ & 0.10 \\ 
\end{tabular}
\vspace*{0.2cm}
\begin{tabular}{ll|ll}
\multicolumn{2}{c|}{$5/2^-_1$}             & \multicolumn{2}{c}{$5/2^-_2$}               \\ \hline
$(0s)^4(0p_{3/2})^2_{21}(0p_{1/2})$ & 0.64 & $(0s)^4(0p_{3/2})^3$                & 0.61  \\
$(0s)^4(0p_{3/2})(0p_{1/2})^2_{10}$ & 0.09 & $(0s)^4(0p_{3/2})^2_{30}(0p_{1/2})$ & 0.12  \\
$(0s)^4(0p_{3/2})^2_{30}(0p_{1/2})$ & 0.06 & $(0s)^4(0p_{3/2})(0p_{1/2})^2_{10}$ & 0.05  \\ 
\end{tabular}
\vspace*{0.2cm}
\begin{tabular}{ll}
\multicolumn{2}{c}{$5/2^-_3$}             \\ \hline
$(0s)^4(0p_{3/2})^2_{30}(0p_{1/2})$ & 0.60 \\ 
$(0s)^4(0p_{3/2})^3$                & 0.11 \\
$(0s)^4(0p_{3/2})^2_{21}(0p_{1/2})$ & 0.06 \\
\end{tabular}
\vspace*{0.2cm}
\begin{tabular}{ll|ll}
\multicolumn{2}{c|}{$7/2^-_1$}             & \multicolumn{2}{c}{$7/2^-_2$}               \\ \hline
$(0s)^4(0p_{3/2})^3$                & 0.53 & $(0s)^4(0p_{3/2})^2_{30}(0p_{1/2})$ & 0.53  \\
$(0s)^4(0p_{3/2})^2_{30}(0p_{1/2})$ & 0.29 & $(0s)^4(0p_{3/2})^3$                & 0.29  \\
\end{tabular}
\end{ruledtabular}
\end{table}


For $^8$Li, various energy components are shown in Table \ref{ham8} measured from the $^4$He components.
The dominant configuration are listed in Table \ref{conf8}.
In $^8$Li, the $2^+$ state is the ground state and has the largest tensor contribution in this nucleus.
This property can be understood from the tensor correlation;
The $2^+$ state dominantly has the $(0p_{3/2})^4$ configuration instead of using the $0p_{1/2}$ orbit.
This allows the selected excitation from the $0s$-orbit to the $0p_{1/2}$ one in the $^4$He part and increase the tensor correlation in this state relatively in comparison with the other states of $^8$Li.
The $3^+$ state also show the similar result of the larger tensor contribution.

\begin{table}[t]
\caption{Various energy components in $^8$Li measured from those of the $^4$He ground state. Units are in MeV.}
\begin{tabular}{ccccccc}
\noalign{\hrule height 0.5pt}
$J^\pi$  & Kinetic& Central  & Tensor   & $LS$     \\
\noalign{\hrule height 0.5pt}
$0^+$    &  62.47 & $-34.15$ & $-14.30$ & $-1.07$  \\
$1^+_1$  &  64.04 & $-34.21$ & $-14.71$ & $-3.51$  \\
$1^+_2$  &  62.35 & $-33.64$ & $-13.11$ & $-2.18$  \\
$1^+_3$  &  61.69 & $-31.86$ & $-13.17$ & $-1.89$  \\
$1^+_4$  &  59.58 & $-29.30$ & $-11.59$ & $-2.88$  \\
$2^+_1$  &  72.08 & $-38.69$ & $-19.30$ & $-4.69$  \\
$2^+_2$  &  69.67 & $-36.53$ & $-16.41$ & $-3.10$  \\
$2^+_3$  &  69.23 & $-37.16$ & $-14.08$ & $-3.71$  \\
$3^+$    &  70.02 & $-36.65$ & $-15.68$ & $-4.86$  \\
$4^+$    &  62.91 & $-31.11$ & $-12.43$ & $-3.32$  \\
\noalign{\hrule height 0.5pt}
\end{tabular}
\label{ham8}
\end{table}

\begin{table}[t]
\centering
\caption{Dominant configurations of $^8$Li($J^\pi$($T$)) with their squared amplitudes $(A^J_k)^2$.
Two subscripts are the spin-isospin quantum numbers.}
\label{conf8}
\begin{ruledtabular}
\begin{tabular}{ll|ll}
\multicolumn{2}{c|}{$0^+ $}                        & \multicolumn{2}{c}{$4^+$}                     \\ \hline
$(0s)^4(0p_{3/2})^2_{01}(0p_{1/2})^2_{01}$  & 0.61 & $(0s)^4(0p_{3/2})^4$                   & 0.81 \\
$(0s)^4(0p_{3/2})^3 (0p_{1/2})$             & 0.20 &                                        &      \\
\end{tabular}
\vspace*{0.2cm}
\begin{tabular}{ll|ll}
\multicolumn{2}{c|}{$1^+_1$}                    & \multicolumn{2}{c}{$1^+_2$}                      \\ \hline
$(0s)^4(0p_{3/2})^3_{3/2,3/2}(0p_{1/2})$ & 0.32 & $(0s)^4(0p_{3/2})^3_{3/2,1/2}(0p_{1/2})$  & 0.50 \\
$(0s)^4(0p_{3/2})^4$                     & 0.25 & $(0s)^4(0p_{3/2})^3_{3/2,3/2}(0p_{1/2})$  & 0.11 \\
$(0s)^4(0p_{3/2})^3_{3/2,1/2}(0p_{1/2})$ & 0.14 & $(0s)^4(0p_{3/2})^2_{10}(0p_{1/2})^2_{01}$& 0.09 \\ 
\end{tabular}
\vspace*{0.2cm}
\begin{tabular}{ll|ll}
\multicolumn{2}{c|}{$1^+_3$}                      & \multicolumn{2}{c}{$1^+_4$}                        \\ \hline
$(0s)^4(0p_{3/2})^2_{01}(0p_{1/2})^2_{10}$ & 0.27 & $(0s)^4(0p_{3/2})^4$                       & 0.27  \\
$(0s)^4(0p_{3/2})^3_{3/2,1/2}(0p_{1/2})$   & 0.23 & $(0s)^4(0p_{3/2})^2_{01}(0p_{1/2})^2_{10}$ & 0.23  \\
$(0s)^4(0p_{3/2})^3_{3/2,3/2}(0p_{1/2})$   & 0.14 & $(0s)^4(0p_{3/2})^3_{3/2,3/2}(0p_{1/2})$   & 0.16  \\ 
$(0s)^4(0p_{3/2})^2_{10}(0p_{1/2})^2_{01}$ & 0.13 & $(0s)^4(0p_{3/2})^2_{10}(0p_{1/2})^2_{01}$ & 0.06  \\ 
\end{tabular}
\vspace*{0.2cm}
\begin{tabular}{ll|ll}
\multicolumn{2}{c|}{$2^+_1$}                      & \multicolumn{2}{c}{$2^+_2$}                       \\ \hline
$(0s)^4(0p_{3/2})^4$                       & 0.41 & $(0s)^4(0p_{3/2})^3_{3/2,3/2}(0p_{1/2})$   & 0.35 \\
$(0s)^4(0p_{3/2})^3_{3/2,3/2}(0p_{1/2})$   & 0.15 & $(0s)^4(0p_{3/2})^3_{3/2,1/2}(0p_{1/2})$   & 0.28 \\
$(0s)^4(0p_{3/2})^2_{21}(0p_{1/2})^2_{01}$ & 0.14 & $(0s)^4(0p_{3/2})^2_{21}(0p_{1/2})^2_{10}$ & 0.08 \\ 
\end{tabular}
\vspace*{0.2cm}
\begin{tabular}{ll|ll}
\multicolumn{2}{c|}{$2^+_3$}                      & \multicolumn{2}{c}{$3^+$}                         \\ \hline
$(0s)^4(0p_{3/2})^3_{3/2,1/2}(0p_{1/2})$   & 0.39 & $(0s)^4(0p_{3/2})^4$                       & 0.53 \\
$(0s)^4(0p_{3/2})^3_{3/2,3/2}(0p_{1/2})$   & 0.20 & $(0s)^4(0p_{3/2})^2_{30}(0p_{1/2})^2_{01}$ & 0.15 \\
$(0s)^4(0p_{3/2})^4$                       & 0.16 & $(0s)^4(0p_{3/2})^3_{5/2,1/2}(0p_{1/2})$   & 0.12 \\ 
\end{tabular}
\end{ruledtabular}
\end{table}

For $^9$Li, various energy components are shown in Table \ref{ham9} and the dominant configuration are listed in Table \ref{conf9}.
This nucleus is important in relation with the structures of more neutron-rich systems of $^{10}$Li and $^{11}$Li and the breaking of the $0p$-shell magicity in that region.
From Table \ref{conf9}, in the ground $3/2^-_1$ state,
the nucleons in the $0p$-shell are dominated by the $(0p_{3/2})^5$ configuration which is a sub-closed configuration for neutron part.
Two neutrons among them can be excited into $0p_{1/2}$ orbit with about 26\%, which is the neutron pairing correlation in the $0p$-shell.
This situation of the $3/2^-_1$ state can allow the excitation of $pn$-pair from $0s$-shell, which emerges the tensor correlation in $^9$Li.
In our previous work, we have used this idea to explain the breaking of the neutron magicity in $N=8$ in $^{11}$Li and also the inversion phenomena of $p$-$sd$ shells in $^{10}$Li.
In that work, the $2p2h$ excitations from the neutron sub-closed configuration of $^9$Li is taken into account 
as the $pn$ and $nn$ pair excitations.
It is also found that the main component of the excited $3/2^-_2$ state can correspond to the pairing excited one with respect to the ground $3/2^-_1$ state.

\begin{table}[t]
\caption{Various energy components in $^9$Li measured from those of the $^4$He ground state. Units are in MeV.}
\begin{tabular}{ccccccc}
\noalign{\hrule height 0.5pt}
$J^\pi$         & Kinetic& Central  & Tensor   & $LS$     \\
\noalign{\hrule height 0.5pt}
1/2$^-$         & 90.78 & $-52.00$ & $-20.08$  & $-4.52$ \\
3/2$^-_1$       & 91.49 & $-53.35$ & $-18.62$  & $-5.72$ \\
3/2$^-_2$       & 89.76 & $-50.40$ & $-19.24$  & $-3.15$ \\
5/2$^-$         & 90.66 & $-51.47$ & $-18.42$  & $-4.20$ \\
7/2$^-$         & 90.05 & $-49.00$ & $-17.68$  & $-5.19$ \\
\noalign{\hrule height 0.5pt}
\end{tabular}
\label{ham9}
\end{table}

\begin{table}[t]
\centering
\caption{Dominant configurations of $^9$Li($J^\pi$($T$)) with their squared amplitudes $(A^J_k)^2$.
Two subscripts are the spin-isospin quantum numbers.}
\label{conf9}
\begin{ruledtabular}
\begin{tabular}{llc}
\multicolumn{2}{c}{$1/2^-$}                    \\ \hline
$(0s)^4(0p_{3/2})^4_{02}(0p_{1/2})$    & 0.67  \\
$(0s)^4(0p_{3/2})^2_{01}(0p_{1/2})^3$  & 0.11  \\
\end{tabular}
\vspace*{0.2cm}
\begin{tabular}{ll}
\multicolumn{2}{c}{$3/2^-_1$}                          \\ \hline
$(0s)^4(0p_{3/2})^5$                            & 0.46 \\
$(0s)^4(0p_{3/2})^3_{3/2,1/2}(0p_{1/2})^2_{01}$ & 0.19 \\
$(0s)^4(0p_{3/2})^3_{3/2,3/2}(0p_{1/2})^2_{01}$ & 0.07 \\ 
\end{tabular}
\vspace*{0.2cm}
\begin{tabular}{ll}
\multicolumn{2}{c}{$3/2^-_2$}                          \\ \hline
$(0s)^4(0p_{3/2})^3_{3/2,3/2}(0p_{1/2})^2_{01}$ & 0.38 \\
$(0s)^4(0p_{3/2})^4_{11}(0p_{1/2})$             & 0.27 \\
$(0s)^4(0p_{3/2})^3_{3/2,3/2}(0p_{1/2})^2_{10}$ & 0.08 \\ 
\end{tabular}
\vspace*{0.2cm}
\begin{tabular}{ll|ll}
\multicolumn{2}{c|}{$5/2^-$}                           & \multicolumn{2}{c}{$7/2^-$}                        \\ \hline
$(0s)^4(0p_{3/2})^4_{21}(0p_{1/2})$             & 0.57 & $(0s)^4(0p_{3/2})^4_{31}(0p_{1/2})$        & 0.80  \\
$(0s)^4(0p_{3/2})^3_{5/2,1/2}(0p_{1/2})^2_{01}$ & 0.13 &                                            &       \\
\end{tabular}
\end{ruledtabular}
\end{table}

Seeing the whole structures from $^6$Li to $^9$Li, it is found that 
their ground states possesses the largest tensor energies, in which 
the configurations of $0p$-shell are constructed to activate the two-kinds of excitations by the tensor force;
one is from $0s$-orbit to the $p_{1/2}$ orbit and the other is from $0p$-shell to higher shell such as $sd$-shell.
In their ground states, the $0p_{1/2}$ component in their most dominant configurations is included only in $^6$Li.
Other heavier systems contain the $(0p_{3/2})^{A-4}$ configuration largely, which is the $jj$-like one, to increase the tensor correlation from the $0s$ orbit to the $0p_{1/2}$ orbit.
From this result, the $^6$Li nucleus is a specific one and this is considered to be related to the $\alpha$+$d$ component in this nucleus, as was mentioned.

\section{Summary}\label{sec:summary}

We have developed a method to describe nuclei with bare $NN$ interaction on the basis of the tensor optimized shell model with the unitary correlation operator method, TOSM+UCOM. 
We have treated the tensor force in terms of TOSM, in which $2p2h$ states are fully optimized to describe the deuteron-like tensor correlation. 
The short-range repulsion in the $NN$ interaction is treated by using the central correlation part of UCOM. 
We have shown the reliability of TOSM+UCOM using the AV8$^\prime$ interaction to investigate the structures of the Li isotopes. 
It is found that the excitation energy spectra are found quite consistent with the experimental spectra.
When we employ the effective Minnesota interaction consisting of only the central and $LS$ parts, the excitation energy spectra show quite a large deviation from experiment. 

It has been found that $^4$He contains a relatively large amount of the $pn$ pair in the $p_{1/2}$ orbit due to the tensor force. 
This characteristics of the tensor force produces the state-dependence in the Li isotopes.
In $^5$Li, the $3/2^-$ state gains more tensor energy than the $1/2^-$ case. 
The enhancement of the kinetic energy is also observed in the $3/2^-$ state of $^5$Li, which is brought by the tensor force. 
As a result, tensor force dynamically produces the state dependence in $^5$Li and contributes to the $LS$ splitting energy in $^5$Li. 
In $^6$Li, the tensor force also makes the large mixing of the spin-triplet component in the configuration of the last $pn$ pair in the ground state, which shows the $LS$ coupling scheme.
This component increases the tensor energy and is related to the $\alpha$+$d$ clustering configuration in $^6$Li. 
This characteristics cannot be seen in the Minnesota interaction without the tensor force, which shows rather the conventional $jj$-coupling scheme. 
In $^{7,8,9}$Li, the $jj$-like configurations, which include more $0p_{3/2}$ states than the $0p_{1/2}$ states, can gain the tensor energy, 
because of the allowance of the $pn$-pair excitation from the $0s$-orbit to $0p_{1/2}$ orbit in the $^4$He core part. 
As a conclusion of the roles of the tensor force in the Li isotopes, the tensor energy depends on the configurations of nucleons occupied mainly in $0p$-shell in each state. 
This property is related with the amount of the excitation of $pn$ pair from $0s$-shell by the tensor force.

The amount of the high momentum component also depends on the tensor energy in each state.
Observation of the high momentum component experimentally in finite nuclei is desired in order to confirm the existence of the strong tensor correlation \cite{subedi08, tanihata10}.

The tensor force would play important roles on various physical quantities such as charge radii, quadrupole and magnetic moments\cite{noertershauser11} in addition to the energy spectra.
In our previous study of $^{11}$Li, we discuss those quantities of $^{11}$Li in detail \cite{myo07_11,myo08}.
It is interesting to discuss the effect of the tensor force on these quantities in other Li isotopes using the TOSM+UCOM method.

\section*{Acknowledgments}
We thank Professor Hisashi Horiuchi for fruitful discussions and continuous encouragement.
This work was supported by a Grant-in-Aid for Young Scientists from the Japan Society for the Promotion of Science (No. 21740194) and also JSPS (No. 21540267).
Numerical calculations were performed on a computer system at RCNP, Osaka University.

\def\JL#1#2#3#4{ {{\rm #1}} \textbf{#2}, #4 (#3)}  
\nc{\PR}[3]     {\JL{Phys. Rev.}{#1}{#2}{#3}}
\nc{\PRC}[3]    {\JL{Phys. Rev.~C}{#1}{#2}{#3}}
\nc{\PRA}[3]    {\JL{Phys. Rev.~A}{#1}{#2}{#3}}
\nc{\PRL}[3]    {\JL{Phys. Rev. Lett.}{#1}{#2}{#3}}
\nc{\NP}[3]     {\JL{Nucl. Phys.}{#1}{#2}{#3}}
\nc{\NPA}[3]    {\JL{Nucl. Phys.}{A#1}{#2}{#3}}
\nc{\PL}[3]     {\JL{Phys. Lett.}{#1}{#2}{#3}}
\nc{\PLB}[3]    {\JL{Phys. Lett.~B}{#1}{#2}{#3}}
\nc{\PTP}[3]    {\JL{Prog. Theor. Phys.}{#1}{#2}{#3}}
\nc{\PTPS}[3]   {\JL{Prog. Theor. Phys. Suppl.}{#1}{#2}{#3}}
\nc{\PRep}[3]   {\JL{Phys. Rep.}{#1}{#2}{#3}}
\nc{\AP}[3]     {\JL{Ann. Phys.}{#1}{#2}{#3}}
\nc{\JP}[3]     {\JL{J. of Phys.}{#1}{#2}{#3}}
\nc{\andvol}[3] {{\it ibid.}\JL{}{#1}{#2}{#3}}
\nc{\PPNP}[3]   {\JL{Prog. Part. Nucl. Phys.}{#1}{#2}{#3}}
\nc{\FBS}[3]   {\JL{Few Body Syst.}{#1}{#2}{#3}}

\end{document}